\pdfoutput=1

\documentclass[aps,pre,reprint,showpacs,amsmath,amssymb,final,letterpaper]{revtex4-1}

\usepackage{wasysym}
\usepackage{graphicx}
\usepackage{subfigure}
\usepackage{epstopdf}

\begin{document}


\title{Adaptive networks: coevolution of disease and topology}

\author{Vincent Marceau}
\author{Pierre-Andr{\'e} No\"el}
\author{Laurent H{\'e}bert-Dufresne}
\author{Antoine Allard}
\author{Louis J. Dub{\'e}}
\affiliation{D\'epartement de physique, de g\'enie physique, et d'optique, Universit\'e Laval, Qu\'ebec, Qu{\'e}bec, Canada G1V 0A6}

\date{\today}

\begin{abstract}
Adaptive networks have been recently introduced in the context of disease propagation on complex networks. They account for the mutual interaction between the network topology and the states of the nodes. Until now, existing models have been analyzed using low complexity analytic formalisms, revealing nevertheless some novel dynamical features. However, current methods have failed to reproduce with accuracy the simultaneous time evolution of the disease and the underlying network topology. In the framework of the adaptive SIS model of Gross \emph{et al.}~[Phys. Rev. Lett. \textbf{96}, 208701 (2006)], we introduce an improved compartmental formalism able to handle this coevolutionary task successfully. With this approach, we analyze the interplay and outcomes of both dynamical elements, \emph{process} and \emph{structure}, on adaptive networks featuring different degree distributions at the initial stage.  
\end{abstract}


\maketitle


\section{Introduction \label{sec:intro}}

The vast majority of network-based models of disease propagation rely on the paradigm of \emph{static networks}~\cite{keeling05_jrsi,meyers07_bams}. In this framework, the assumption is made that the time scale which characterizes the disease propagation is much shorter than the time scale with which the network structure changes. In contrast to static networks, some researchers have investigated the phenomenon of disease propagation on dynamically evolving networks, and have revealed new perspectives on the effects of concurrent or casual partnerships~\cite{bauch02_jmb,eames04_mbs,doherty06std,Morris2009ajph}, contact mixing~\cite{fefferman07_pre,volz07_prsb,volz09_jrsi}, and demographic changes~\cite{kamp2009arxiv}. In these models, however, the rules which govern the evolution of the network are independent of what happens on the network. Mutual interactions between the network topology and the states of the nodes are not taken into account. 
 
Recently, interest has grown for a new class of networks known under the name of \emph{adaptive networks}~\cite{gross08_jrsi,gross2009}. They are characterized by the existence of a feedback loop between the \emph{dynamics on the network} and the \emph{dynamics of the network}. Among other applications, adaptive networks have been introduced in the study of contact processes, such as the study of opinion formation~\cite{gil06_pla,holme06_pre,zanette06_physd,kozma08_pre,nardini08_prl,vazquez08_prl,biely09_epjb,iniguez09_pre,mandra09_pre,sobkowicz09_ijmpc} and epidemic spreading~\cite{gross06_prl,gross08_epl,risau-gusman09_jtb,shaw08_pre,zanette08_jbp}. In epidemiological settings, the main idea behind models featuring adaptive networks is that individuals may change their behavior under the threath of an emerging disease~\cite{schwartz10_physics}. For example, healthy individuals may try to reduce their chance of catching the disease by adaptively replacing their contacts with infectious individuals by contacts with noninfectious ones. This may significantly alter the structure of the contact network, thus influencing the way the disease will spread. 

Being an emerging field of research, the study of contact processes on adaptive networks still lacks strong theoretical foundations. Until now, the analytic treatment of epidemic models on adaptive networks has been limited to low order moment-closure approximations~\cite{gross06_prl,risau-gusman09_jtb,shaw08_pre,zanette08_jbp}. Despite their low complexity, these approaches were able to predict novel dynamical features, such as bistability, hysteresis and first order transitions. However, their simple design does not allow to make accurate predictions about the time evolution of the system. An integrated analytic formalism able to account for the complete time evolution of both dynamical elements, i.e. the spreading disease and the evolving network topology, is still lacking. 

In this paper, we present an analytic approach with the purpose of filling this important gap. Using for its simplicity the epidemic model of Gross \emph{et al.}~\cite{gross06_prl} as the basic  framework, we develop an improved compartmental formalism in which nodes are categorized not only by their state of infectiousness, but also by the state of their neighbors. With this tool, we study the interplay and outcomes of disease and topology on adaptive networks with various initial configurations. Even if we restrict ourselves to one particular model, the approach presented here is quite general and could easily be applied to the study of other contact processes on adaptive networks.

The paper is organized as follows. In Sec.~\ref{sec:model}, we recall the model of Gross \emph{et al.}, and introduce our formalism in Sec.~\ref{sec:formalism}. Analytic predictions are compared with the results obtained from numerical simulations in Secs.~\ref{sec:timeevol} and~\ref{sec:stationary}. More precisely, we concentrate on the time evolution of the system in Sec.~\ref{sec:timeevol}, while its stationary states are investigated in Sec.~\ref{sec:stationary}. Finally, we give further remarks on the endemic stationary state of the system in Sec.~\ref{sec:endemic}, and summarize our conclusions in Sec.~\ref{sec:conclusion}.

\section{SIS dynamics on adaptive networks \label{sec:model}}

We will focus on a simple epidemic model on adaptive networks introduced by Gross \emph{et al.}~\cite{gross06_prl}. We consider a random dynamical network consisting of a fixed number of nodes $N$ and undirected links $M=\langle k \rangle N/2$, where $\langle k \rangle$ is defined as the average degree (number of links per node) of the network. The nodes of the network represent the individuals of a given population, while the links stand for potential disease-causing contacts between pairs of individuals. Two nodes are said to be neighbors if they are joined by a link. Neither can a node be linked to itself (no self-loops) nor share more than one link with another node (no repeated links). The set of probabilities $\{p_k(t)\}$ that a node chosen at random at time $t$ is of degree $k$, called the degree distribution, characterizes the topology of the network at this particular time. The mean degree of a network corresponds to the first moment of its degree distribution, $\langle k \rangle = \sum_k kp_k(t) = 2M/N$.

We consider a case of Susceptible-Infectious-Susceptible (SIS) dynamics. At any time, each node is in a specific state, either \emph{susceptible} (S) or \emph{infectious} (I). Infectious individuals contaminate their susceptible neighbors at rate $\beta$, while they recover and become susceptible again at rate $\alpha$. The coupling between disease and topology is implemented by adding an adaptive rewiring rule. Susceptible individuals are allowed to replace at rate $\gamma$ their infectious neighbors by individuals chosen at random in the susceptible population. These rules guarantee that $N$ and $M$ remain constant over time. Even if the system contains three dynamical parameters, its behavior is characterized by two independent dimensionless ratios, e.g. $\beta/\alpha$ and $\gamma/\alpha$, since time can always be rescaled according to one parameter.

To perform Monte-Carlo simulations of epidemic propagation on a network, one requires an explicit knowledge of the network structure. Our networks are generated according to the following algorithm~\cite{newman02_pre}. We first generate a random degree sequence $\{ k_i \}$ of length $N$ subjected to the initial degree distribution specified by $\{p_k(0)\}$. In this process, we make sure that $\sum_i k_i$ is even since each link consists of two ``stubs''. For each node $i$, a node with $k_i$ stubs is produced, then pairs of unconnected stubs are randomly chosen and connected together until all unconnected stubs are exhausted. Afterwards, we test for the presence of self-loops and repeated links. All faulty links are removed by randomly choosing a pair of connected stubs and rewiring them to the former stubs.

Monte-Carlo simulations of SIS dynamics on adaptive networks are carried out using discrete time steps of length $\Delta t$. At each step, the recovery, infection and rewiring events are tested with probabilities $\alpha \Delta t$, $\beta \Delta t$, and $\gamma \Delta t$ respectively. Self-loops and repeated links are explicitly forbidden during the rewiring process. All simulations start with the random infection of a fraction $\epsilon$ of the individuals in the network. We use the parameters $\Delta t = 0.1$ and $N=25000$ in all simulations. The recovery rate $\alpha=0.005$ is used unless explicitly noted.

In what follows, we perform simulations on adaptive networks featuring different initial degree distributions. The first distribution to be used is the delta distribution,
\begin{align}
p_k^{DR} = \delta_{k,k_0} \ ,
\label{eq:drdist}
\end{align}
which produces a \emph{degree-regular (DR)} network where each node has the same degree $k_0$. The second type of distribution considered is the Poisson distribution,
\begin{align}
p_k^{P} = \frac{z^k e^{-z}}{k!} \ ,
\label{eq:pdist}
\end{align}
which corresponds, in the limit $N\gg1$, to networks in which the presence of a link between two nodes is governed by the same probability, independent of the links already present in the network. We will refer to them as \emph{Poisson (P)} networks, and their mean degree is given by $\langle k \rangle^{P} = z$. Finally, we will also use a truncated power law distribution,
\begin{align}
p_k^{PL} & = \begin{cases}
                \ \frac{1}{C}k^{-\tau}\ ,      & 0 < k \leq k_c \\
                \ 0\ ,     & k > k_c
           \end{cases} \ , 
           \label{eq:pldist}
\end{align}
where $\tau>0$ and $C=\sum_{k=1}^{k_c}k^{-\tau}$ so that the distribution is properly normalized. This produces \emph{power law distributed (PL)} networks, where highly connected hubs and individuals with few connections coexist. The mean degree of networks generated by Eq. \eqref{eq:pldist} is given by $\langle k \rangle^{PL} = C^{-1} \sum_{k=1}^{k_c} k^{1-\tau}$. To obtain a network with $\langle k \rangle^{PL}=2$, we use $\tau=2.16104$ and $k_c=20$.

\section{Improved compartmental formalism \label{sec:formalism}}

In order to describe the complete time evolution of the model defined in the last section, we introduce an improved compartmental formalism in the spirit of the formalism presented in the Appendix of No\"el \emph{et al.}~\cite{noel09_pre}. 

\subsection{Dynamical equations \label{sec:dynaeq}}

Let $S_{kl}(t)$ and $I_{kl}(t)$ be the fractions of nodes of \emph{total degree} $k$ and \emph{infectious degree} $l \leq k$ that are respectively susceptible and infectious at time $t$~\footnote{For the sake of readability, we will drop from here on the explicit time dependence of $S_{kl}$, $I_{kl}$, and related quantities. Stationary values will be marked with an asterisk $(^*)$.}. Here by total degree we mean the total number of links that belong to a node, and by infectious degree the number of those links shared with infectious individuals. We define the zeroth order moments of the $S_{kl}$ and $I_{kl}$ distributions by
\begin{align}
S \equiv \sum_{kl}S_{kl} \ \ \textrm{and} \ \ I \equiv \sum_{kl}I_{kl} \ ;
\end{align}
the first order moments by
\begin{align}
\begin{split}
&S_S \equiv \sum_{kl}(k-l)S_{kl} \ , \ S_I \equiv \sum_{kl}lS_{kl} \ , \\
&I_S \equiv \sum_{kl}(k-l)I_{kl} \,  \ \textrm{and} \ \ I_I \equiv \sum_{kl}lI_{kl} \ ;
\end{split}
\end{align}
and the second order moments by
\begin{align}
S_{SI} \equiv \sum_{kl}(k-l)lS_{kl} \ , \ S_{II} \equiv \sum_{kl} l(l-1) S_{kl}  \ , \textrm{ etc.}
\end{align}
Physically, the zeroth order moments correspond to the density of S and I nodes, the first order moments to the density per node of the various types of arcs, and the second order moments to the density per node of the various types of triplets in the network~\footnote{An arc is defined as a \emph{directional} connection between two nodes. Each undirected link in the network thus consists of two directional arcs. A triplet is defined as the union of two directional arcs around the same central node.}.

\begin{figure}[t]
\mbox{
\subfigure[]{\includegraphics[width = .25\columnwidth]{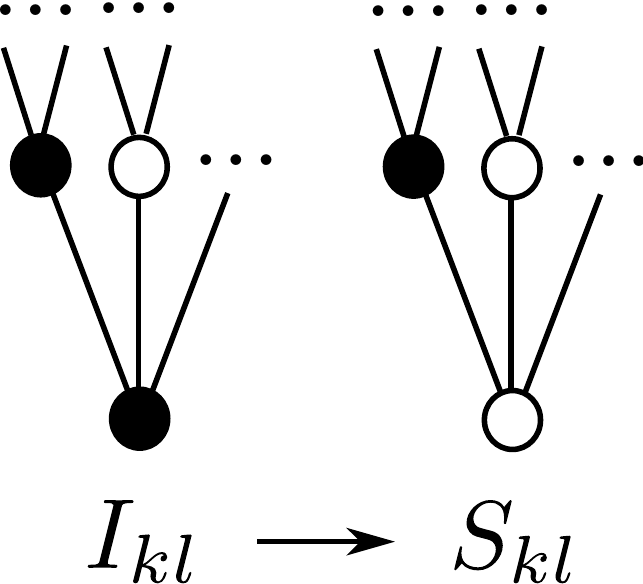} \label{fig:move1}} \ \ \
\subfigure[]{\includegraphics[width = .25\columnwidth]{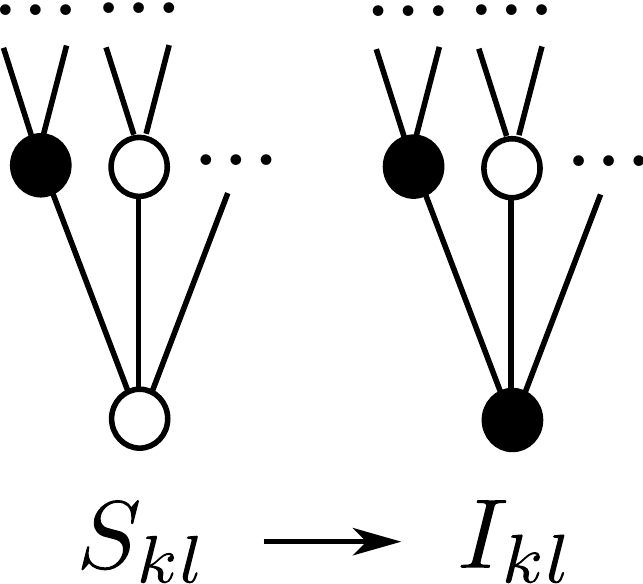} \label{fig:move2}} \ \ \  
\subfigure[]{\includegraphics[width = .25\columnwidth]{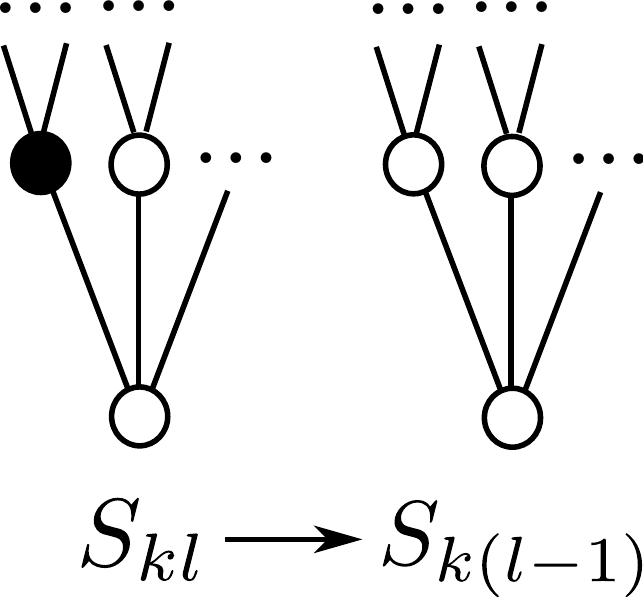} \label{fig:move3}}} \\
\mbox{
\subfigure[]{\includegraphics[width = .25\columnwidth]{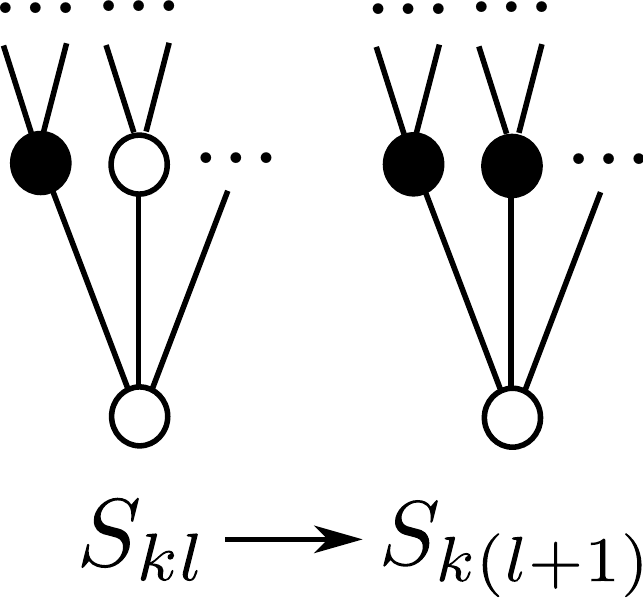} \label{fig:move4}} \ \ \
\subfigure[]{\includegraphics[width = .25\columnwidth]{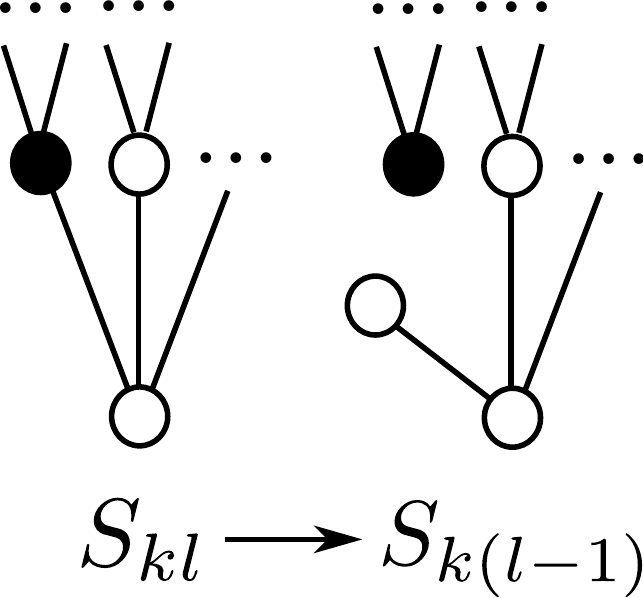} \label{fig:move5}} \ \ \ 
\subfigure[]{\includegraphics[width = .25\columnwidth]{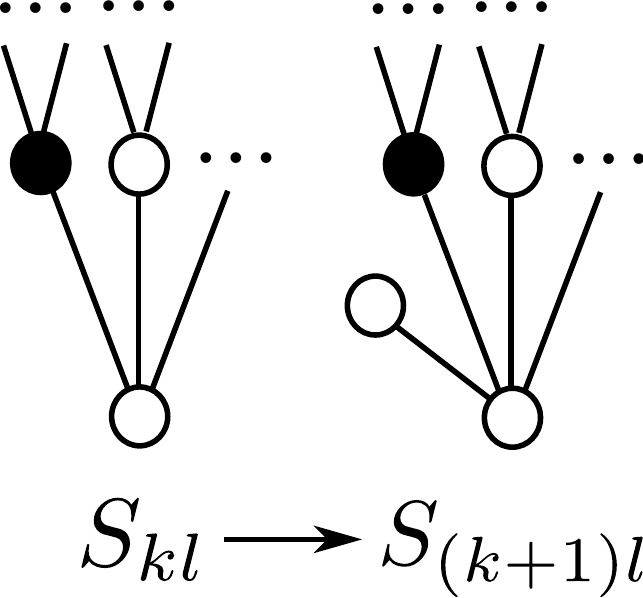} \label{fig:move6}}} 
\caption{Schematic illustration of the events described in the text which result in changing a node from one compartment to another. Susceptible nodes are represented by open symbols (\Circle) and infectious nodes by filled symbols (\CIRCLE). \label{fig:moves}}
\end{figure}

As mentionned in the last section, our model is constrained by two conservation relations, namely the conservation of nodes, 
\begin{align}
S + I = 1\ ,
\label{eq:c1}
\end{align}
and the conservation of links,
\begin{align}
S_S + S_I + I_S + I_I = \langle k \rangle \ ,
\label{eq:c2}
\end{align}
which must hold at any time $t$. Moreover, since the network under consideration is undirected, the density of SI links must always equal the density of IS links. This yields the additional constraint 
\begin{align}
S_I = I_S \ .
\label{eq:c3}
\end{align}

\begin{figure*}[!t]
\mbox{
\subfigure[\ $\beta = 0.06$, $\gamma = 0.02$]{\includegraphics[width = .45\textwidth]{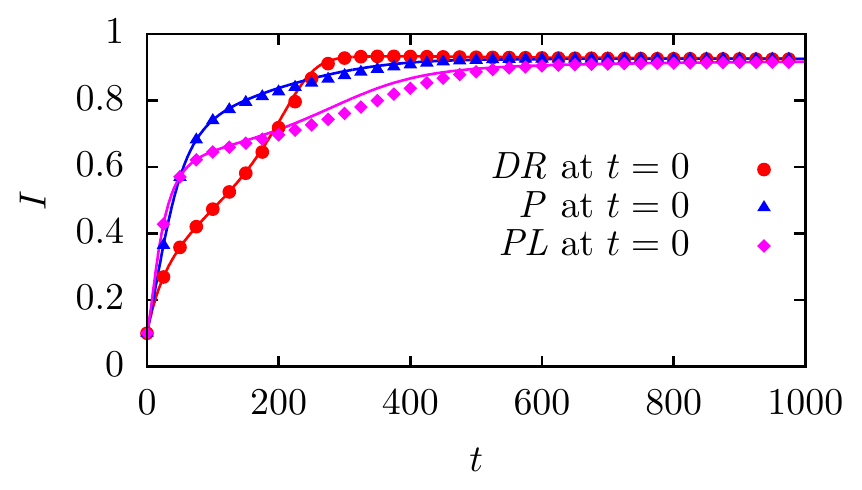} \label{fig:tevol1}} \qquad
\subfigure[\ $\beta = 0.04$, $\gamma = 0.04$]{\includegraphics[width = .45\textwidth]{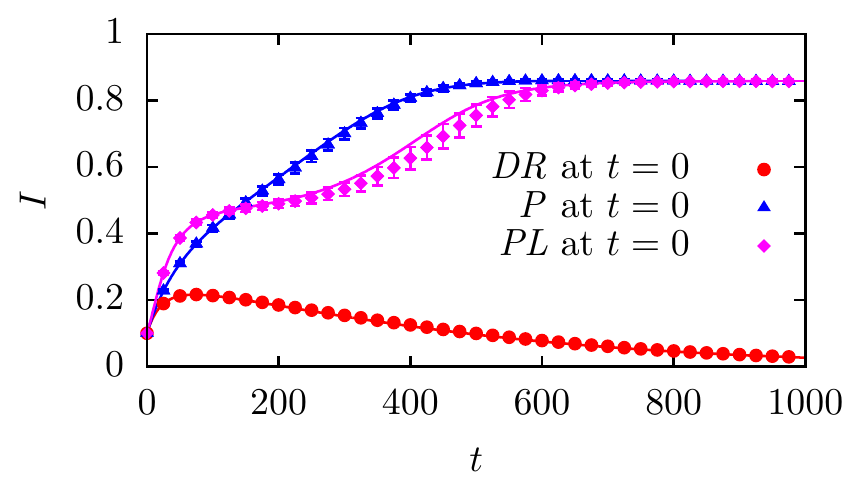} \label{fig:tevol2}}} 
\caption{(Color online) Disease prevalence $I$ against time $t$ on networks featuring the same mean degree $\langle k \rangle=2$ but different initial degree distributions ($p_k^{DR}$, $p_k^P$ and $p_k^{PL}$). Parameters are (a) $\beta = 0.06$, $\gamma = 0.02$ and (b) $\beta = 0.04$, $\gamma = 0.04$. We use $\alpha=0.005$ and $\epsilon=0.1$ in all simulations. Points and error bars (if larger than marker size) correspond to the mean and standard deviation computed over $10000$ Monte-Carlo simulations; solid lines are the predictions of our analytic approach. \label{fig:tevol}}
\end{figure*}

We now derive an ordinary differential equation (ODE) for each compartment of the system. Let us illustrate the reasoning for $S_{kl}$. First, nodes can change compartment according to a change in their own state. Nodes are added to $S_{kl}$ at rate $\alpha I_{kl}$ as nodes from $I_{kl}$ recover and become susceptible again [Fig. \ref{fig:move1}], and are removed from $S_{kl}$ at rate $\beta l S_{kl}$ as they get infected [Fig. \ref{fig:move2}]. 

Second, we have to account for a change in the state of a node's neighbors. Nodes from $S_{kl}$ are transfered to $S_{k(l-1)}$ when one of their infectious neighbors becomes susceptible again [Fig. \ref{fig:move3}], which occurs at rate $\alpha l S_{kl}$. On the opposite, nodes from $S_{kl}$ are moved to $S_{k(l+1)}$ when one of their susceptible neighbors catches the disease [Fig. \ref{fig:move4}]. Since we do not know the exact number of infectious neighbors connected to each of the $(k-l)$ susceptible neighbors of nodes in the $S_{kl}$ compartment, our best option is to take an average over the entire network. In doing so, we make the assumption of \emph{zero degree correlation}: we assume that nodes, beyond the knowledge of the state of infection of each neighbor, are connected at random in the network. Hence, a node in compartment $S_{k'l'}$ with $l'$ infectious neighbors will be reached by following a SS link with a probability equal to $(k'-l')S_{k'l'}/S_{S}$. Thus, $S_{SI}/S_S$ gives the average number of infectious neighbors that a susceptible node reached by following a SS link has. The rate associated to the $S_{kl} \to S_{k(l+1)}$ transition is therefore $\beta (S_{SI}/S_S) (k-l) S_{kl}$. 

Finally, we have to account for the effects of rewiring. Nodes from $S_{kl}$ become labeled as $S_{k(l-1)}$ at rate $\gamma l S_{kl}$ as they break connections with their infectious neighbors [Fig. \ref{fig:move5}]. Moreover, a node from $S_{kl}$ is moved to the compartment $S_{(k+1)l}$ if it is chosen as the ``new neighbor'' in a rewiring event [Fig. \ref{fig:move6}]. Since the strength of rewiring events is $\gamma S_I$ and a node from $S_{kl}$ is randomly chosen with a probability $S_{kl}/S$, this occurs at a rate $\gamma( S_I / S) S_{kl}$. By summing all contributions, we obtain the following ODE governing the time evolution of the $S_{kl}$ compartment: 
\begin{align}
\frac{dS_{kl}}{dt} = & \ \alpha  I_{kl} - \beta l S_{kl} + \alpha \Big[ (l+1)S_{k(l+1)} - lS_{kl} \Big]  \nonumber \\
&+ \beta \dfrac{S_{SI}}{S_S} \Big[ (k-l+1)S_{k(l-1)} - (k-l)S_{kl} \Big] \label{eq:dSkldt} \\
&+ \gamma \Big[ (l+1)S_{k(l+1)} - lS_{kl} \Big] + \gamma\dfrac{S_I}{S}\Big[ S_{(k-1)l} - S_{kl} \Big] \nonumber \ .
\end{align}
A similar reasoning for the $I_{kl}$ compartment yields the following ODE:
\begin{align}
\frac{dI_{kl}}{dt} = &-\alpha I_{kl} + \beta l S_{kl} + \alpha \Big[  (l+1) I_{k(l+1)} - l I_{kl} \Big] \nonumber \\
&+ \beta \left( 1+\dfrac{S_{II}}{S_I}\right) \Big[  (k-l+1)I_{k(l-1)} - (k-l)I_{kl} \Big] \nonumber \\
&+ \gamma \Big[ (k-l+1)I_{(k+1)l} - (k-l)I_{kl} \Big] \ .\label{eq:dIkldt}
\end{align}
It is straightforward to show that the infinite system of ODEs consisting of Eqs. \eqref{eq:dSkldt} and \eqref{eq:dIkldt} satisfies the constraints given by Eqs. \eqref{eq:c1}-\eqref{eq:c3}.

\subsection{Initial conditions \label{sec:inicond}}

In order for the dynamics to be completely specified, we need to write an initial condition for each compartment. In the case where a fraction $\epsilon$ of the nodes is initially infected at random, they are given by
\begin{align}
S_{kl}(0) = (1-\epsilon) p_k(0) \binom{k}{l}\epsilon^l (1-\epsilon)^{k-l} 
\label{eq:Skl0}
\end{align}
and
\begin{align}
I_{kl}(0) = \epsilon p_k(0) \binom{k}{l}\epsilon^l (1-\epsilon)^{k-l} \ .
\label{eq:Ikl0}
\end{align}
Again, we easily verify that this set satisfies Eqs. \eqref{eq:c1}-\eqref{eq:c3}. 

The complete time evolution of the system is obtained by integrating numerically the set of ODEs given by \eqref{eq:dSkldt} and \eqref{eq:dIkldt} truncated at $k_{\textrm{max}}$, together with the initial conditions \eqref{eq:Skl0} and \eqref{eq:Ikl0}. Constraints \eqref{eq:c1}-\eqref{eq:c3} can be used to check the precision of the numerical integration. The complexity of the system of equations is $\mathcal{O}(k_{\textrm{max}}^2)$.

\section{Time evolution: interplay between disease and topology \label{sec:timeevol}}

As stated previously, we initialize the dynamics of the model by infecting a fraction $\epsilon$ of the nodes in the network at random. Afterwards, the states of the nodes and the topology of the network coevolve according to the rules prescribed in Sec.~\ref{sec:model}. In this section, we analyze the time evolution of the system, both from the perspective of the spreading disease and of the evolving network topology. 

A first quantity of interest is the evolution of the disease prevalence, defined as the fraction of infectious individuals at time $t$. According to our previous definitions, it is simply given by $I$. In Fig. \ref{fig:tevol}, we illustrate the evolution of $I$ for networks with different initial topologies, namely \emph{DR}, \emph{P} and \emph{PL} networks featuring a mean degree $\langle k \rangle=2$. In Fig. \ref{fig:tevol1}, all systems reach an endemic steady state where the disease prevalence seems to stabilize at the same value. However, the time evolution of $I$ follows very different patterns depending on the initial configuration of the system. In Fig. \ref{fig:tevol2}, we show that for another set of parameters, the \emph{P} and the \emph{PL} network converge towards an endemic state, whereas the rewiring is sufficiently strong to hinder the initial propagation in the \emph{DR} network and consequently, the system converges to a disease-free state. This indicates that the initial network topology may also influence the global outcome of a particular epidemic scenario.
\begin{figure*}[t]
\mbox{
\subfigure[\ \ \ Low-degree nodes ($k=0$, 1, 2)]{\includegraphics[width = .50\textwidth]{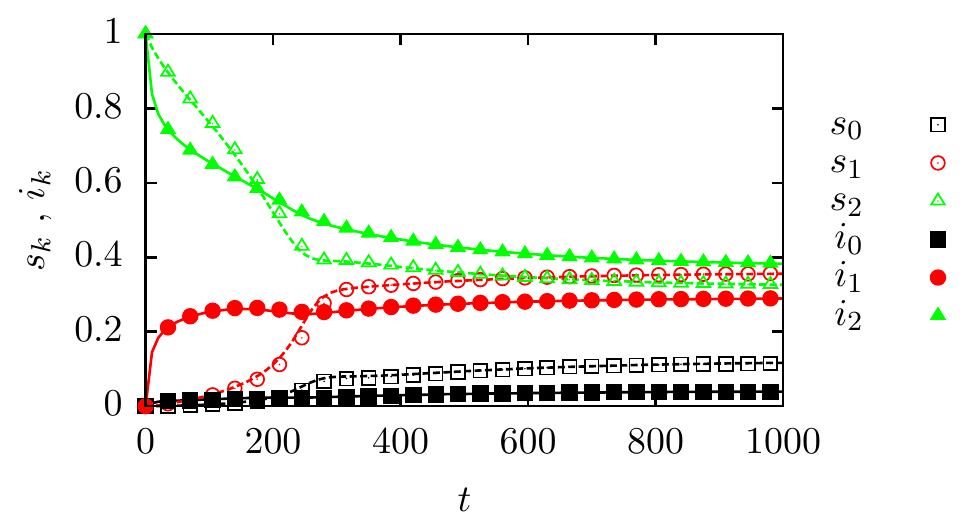} \label{fig:ddevol_low}}
\subfigure[\ \ \ High-degree nodes ($k=3$, 4, 5)]{\includegraphics[width = .50\textwidth]{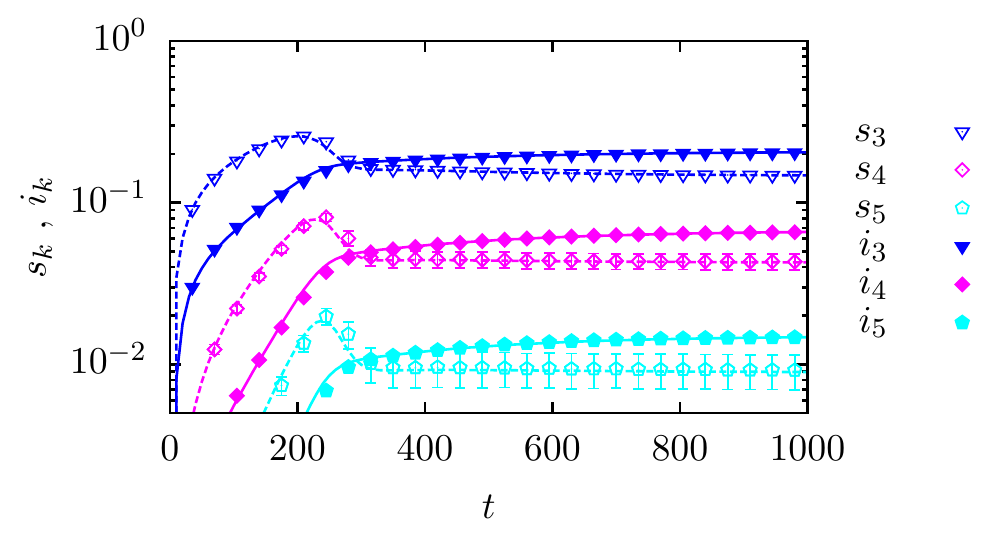} \label{fig:ddevol_high}}}
\caption{(Color online) Degree probability $s_k$ and $i_k$ for susceptible (open symbols, dashed lines) and infectious (filled symbols, solid lines) individuals against time t on an adaptive networks with a \emph{DR} initial degree distribution ($k_0=2$).  The parameters of the system are $\alpha=0.005$, $\beta = 0.06$, $\gamma = 0.02$, and $\epsilon = 0.1$, as in Fig. \ref{fig:tevol1}. Points and error bars (if larger than marker size) correspond to the mean and standard deviation computed over $10000$ Monte-Carlo simulations; curves are the predictions of our analytic approach. \label{fig:ddevol}}
\end{figure*}

As the disease propagates, connections between individuals are being adaptively rewired, which affects the degree distribution of the network. The normalized degree distributions of susceptible and infectious individuals are given in our formalism by  $s_k \equiv \sum_l S_{kl}/S$ and $i_k \equiv \sum_l I_{kl}/I$.

We consider the simplest example of a population initially connected via a \emph{DR} network with $\langle k \rangle=2$. For the same parameters as in Fig. \ref{fig:tevol1}, we show in Fig.~\ref{fig:ddevol} the time evolution of the probabilities $s_k$ and $i_k$ for low-degree ($k=0$, 1, 2) and high-degree ($k=3$, 4, 5) nodes. At $t=0$, all nodes are of degree 2. When the disease starts to propagate, both degree distributions are rapidly modified. The fraction of degree 1 infectious nodes quickly increases, because the susceptible neighbor of a degree 2 node who has just been infected will try to rewire its connection. This result in an increase of degree 3 and higher susceptible nodes. During this first phase of infection, we observe in Fig.~\ref{fig:ddevol_high} that the fraction of high degree susceptible nodes smoothly increases, each degree probability lagging behind the preceding one. Shortly after $t=200$, the fraction of susceptibles that are of degree 3 and higher suddenly drops, while the fraction of them that are either of degree 0 or 1 increases. Since susceptible nodes cannot lose connections, this means that the infection reaches an important number of high degree susceptible nodes that have formed during the initial stage of infection. Afterwards, the topology settles slowly towards its stationary state.

In order to investigate further the interplay between disease and topology, it is useful to look at the evolution of some important observables of the system. First, we consider the fraction of SI links in the network. This quantity, given by $S_I$ (or $I_S$), is directly proportional to the number of new infections at a given moment in time. While $S_I$ is a good measure of the instantaneous \emph{dangerousness} of the situation, it does not yield any information about the potential of the disease to spread further. To quantify the latter effect, we use the \emph{effective branching factor} $\kappa_{IS}^S \equiv S_{SI}/S_I$, which is the average number of susceptible neighbors of a susceptible individual reached by following an IS link. Since $\kappa_{IS}^S$ is related to the degree of correlation between susceptible individuals, we also compute the average fraction $C_{SS}\equiv S_S/(S_S+S_I)$ of connections that susceptible individuals share with other susceptible individuals.

\begin{figure}[t]
\centering
\includegraphics[width = 0.8\columnwidth]{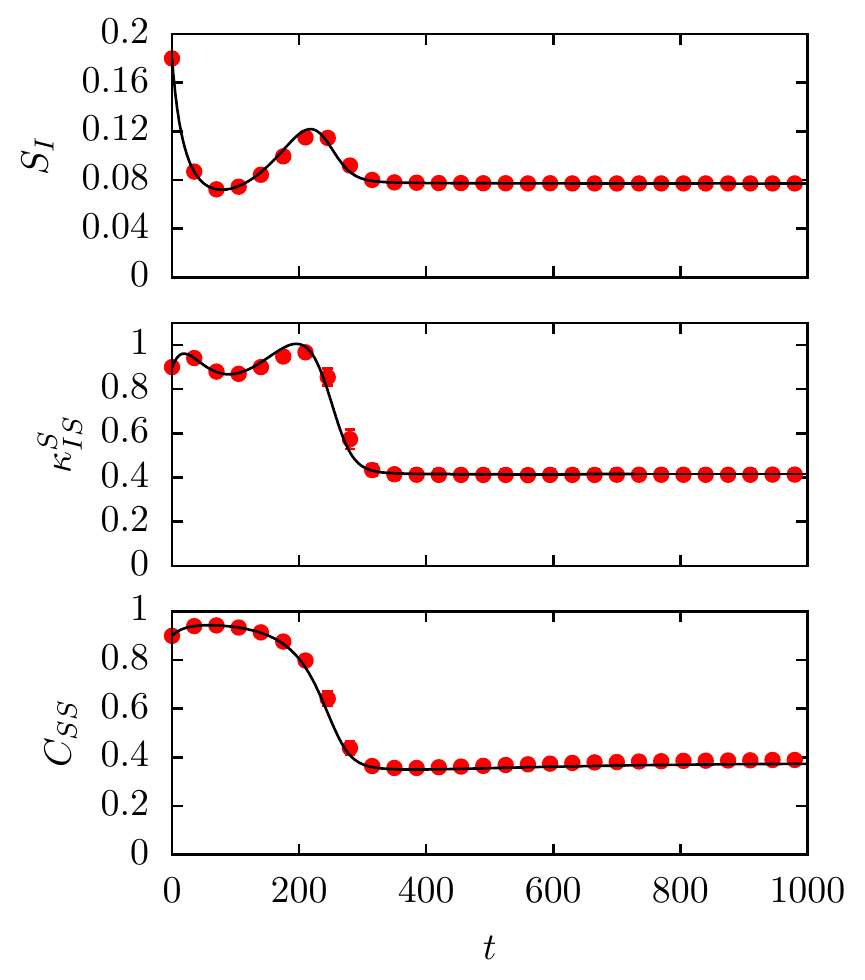} 
\caption{(Color online) Time evolution of the fraction of SI links $S_I$, the effective branching factor $\kappa_{IS}^S$, and the average number of connections that susceptible nodes share with other susceptibles $C_{SS}$ in a system with an adaptive networks featuring a \emph{DR} initial degree distribution ($k_0=2$).  The parameters of the system are $\alpha=0.005$, $\beta = 0.06$, $\gamma = 0.02$, and $\epsilon = 0.1$. Points and error bars (if larger than marker size) correspond to the mean and standard deviation computed over $10000$ Monte-Carlo simulations; solid lines are the predictions of our analytic approach. \label{fig:topoevol}}
\end{figure}

In Fig.~\ref{fig:topoevol}, the time evolution of $S_I$, $\kappa_{IS}^S$ and $C_{SS}$ is illustrated for the example of the initial \emph{DR} network considered so far. We now put these results in parallel with those obtained on Figs.~\ref{fig:tevol1} and~\ref{fig:ddevol}. In the system under consideration, $\beta$ and $\gamma$ are significantly larger than $\alpha$, and therefore, the infection and rewiring processes dominate at early times. At $t=0$, everyone is of degree 2, which means that infecting a susceptible node does not increase the number of SI links. Hence $S_I$ quickly decreases due to adaptive link rewiring and the propagation speed gradually decreases. As we saw in Fig.~\ref{fig:ddevol}, initial rewiring events result in an increase of high-degree susceptible nodes. Morever, we see in Fig.~\ref{fig:topoevol} that those susceptible nodes form a strongly linked community: near $t=100$, $C_{SS}$ stays as high as 0.9 even while nearly half of the nodes in the network are infected. The initial disease propagation phase is thus characterized by a \emph{topological segregation between susceptible and infectious individuals}.

However, this situation is unstable. There still are some SI links in the network, and a strongly connected community of susceptible nodes is very vulnerable to a rapid infection. The disease eventually invades the susceptible community: near $t=100$, the effective branching factor $\kappa_{IS}^S$ rises again, yielding an increase in the number of SI links which reaches a peak near $t=200$. As the disease propagates in the community of susceptible nodes, $C_{SS}$ exhibits a sharp decrease. To this invasion of the tightly linked community of susceptible nodes correspond the second burst of infection observed in Fig.~\ref{fig:tevol1} and the sudden decrease of high-degree susceptibles observed in Fig.~\ref{fig:ddevol}, both near $t=200$. After this second phase of disease propagation, the system converges smoothly towards a stationary state.

Figures~\ref{fig:tevol},~\ref{fig:ddevol}, and \ref{fig:topoevol} confirm that our formalism is well capable of tracking the time evolution of disease and topology on adaptive networks. Numerical results obtained from Monte-Carlo simulations are in excellent agreement with analytic predictions. We may mention that a good agreement between theory and simulations has also been obtained for the time evolution of degree distributions and topological observables in systems featuring $P$ and $PL$ initial networks with the parameters used in Fig. \ref{fig:tevol1}.

\section{Stationary states \label{sec:stationary}}

\begin{figure*}[!t]
\mbox{
\subfigure[\ $\langle k \rangle = 20$]{\includegraphics[width = .32\textwidth]{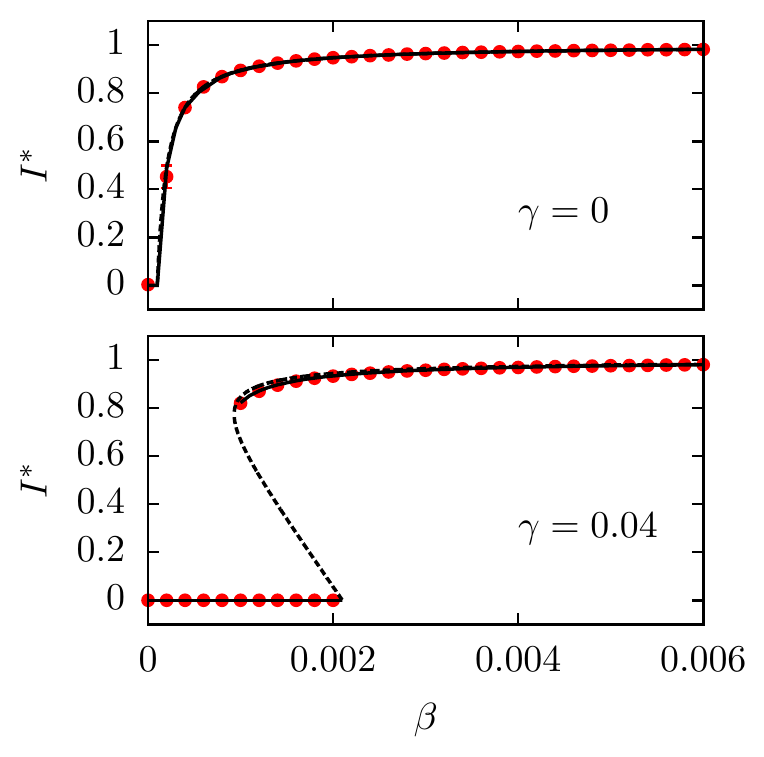} \label{fig:bif_z20}} \
\subfigure[\ $\langle k \rangle = 7$]{\includegraphics[width = .32\textwidth]{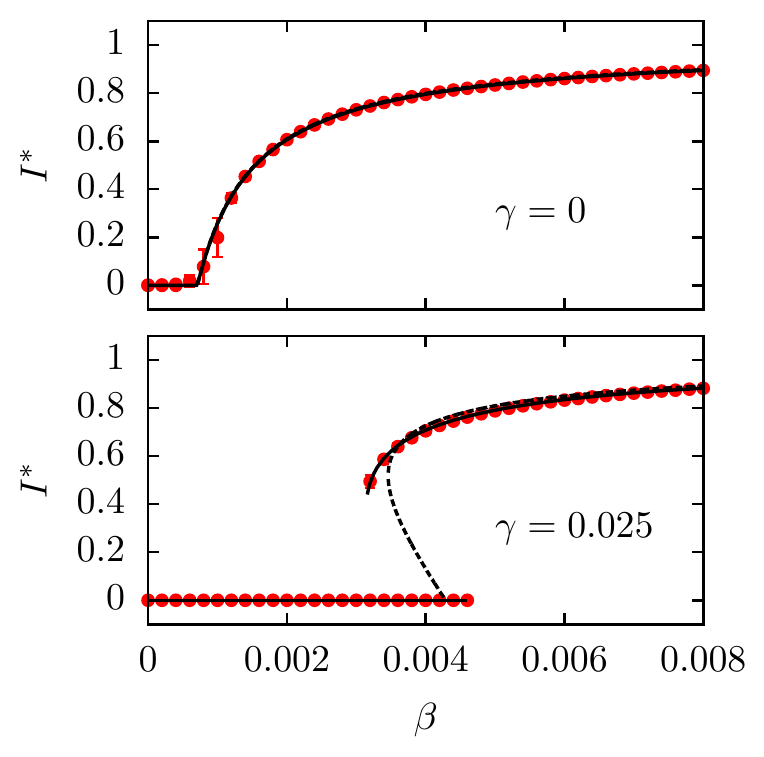} \label{fig:bif_z7}} \  
\subfigure[\ $\langle k \rangle = 2$]{\includegraphics[width = .32\textwidth]{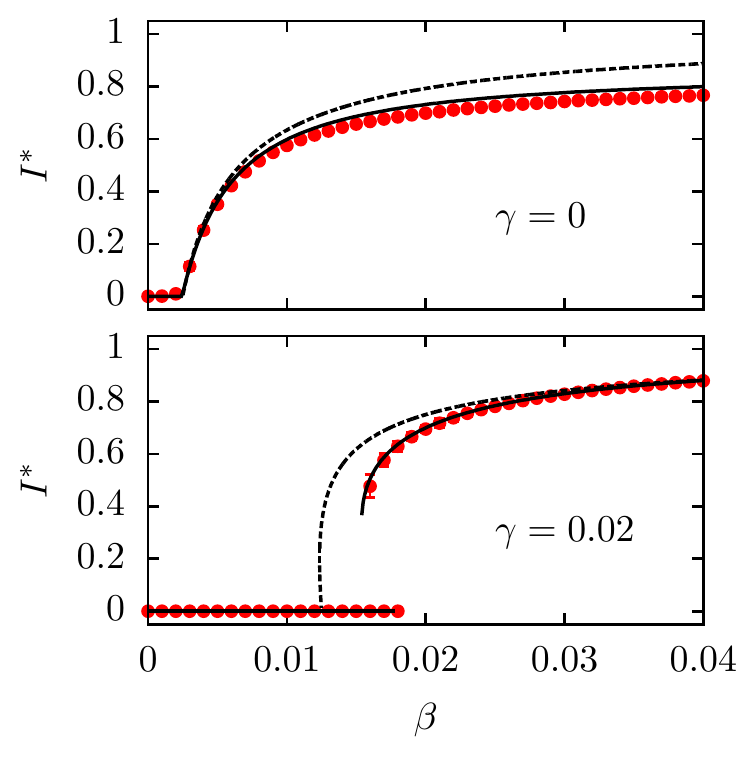} \label{fig:bif_z2}}} 
\caption{(Color Online) Bifurcation diagrams of the stationary disease prevalence $I^*$ versus infection rate $\beta$ on static and adaptive networks with an initial Poisson degree distribution and (a) $\langle k \rangle = 20$ and $\alpha = 0.002$; (b) $\langle k \rangle = 7$ and $\alpha = 0.005$; (c) $\langle k \rangle = 2$ and $\alpha = 0.005$. Solid lines are the predictions of our analytic approach while dashed curves are the predictions of the formalism of Gross \emph{et al.} (stable and unstable manifolds). Points and error bars (if larger than marker size) represent the outcome of Monte-Carlo simulations. To collect statistics in Monte-Carlo simulations, at least 25 simulations were run for each value $\epsilon=0.001$, 0.01, 0.05, 0.99 and 0.999. For each run, the initial transient was discarded, and the prevalence at equilibrium was averaged over at least 10000 time steps. Averages were calculated with the sets of results obtained from this procedure.\label{fig:bif}}
\end{figure*}

After studying the time evolution of the system, we now investigate its stationary states. At first glance, a given epidemic scenario may admit three different outcomes. First, the disease may not be virulent enough to propagate throughout the network, hence the system will converge towards a \emph{disease-free} state, i.e. a \emph{frozen} configuration where all the nodes are susceptible. Second, the disease may reach and maintain a fixed macroscopic prevalence in the population, where the number of new infections equals the number of recoveries at any time. When this \emph{endemic state} is reached, the system is in \emph{active} equilibrium, since infection, recovery and rewiring events continuously occur. Third, we cannot reject the possibility that the disease prevalence may never settle to a constant value, and behave in a periodic, quasiperiodic or even chaotic fashion. In our study, we only reported the first two scenarios, what is consistent with previous studies of this or similar systems~\cite{gross06_prl, zanette08_jbp, shaw08_pre, risau-gusman09_jtb}. The presence of a stable limit cycle in a narrow region of parameter space was theoretically predicted in~\cite{gross06_prl}, and short-lived oscillations were reported in Monte-Carlo simulations of large systems~\footnote{T. Gross, private communication (2010).}. In what follows, we do not report any oscillations. We believe that oscillations could in principle also be found with our analytic framework. However, due to the high complexity of our approach, finding the oscillatory regime would require a systematic investigation of the parameter space beyond the scope of this paper.

\subsection{Bifurcation structure and topology at equilibrium \label{sec:bif}}

In order to study the properties of the system at equilibrium, we first consider the stationary disease prevalence $I^*$. In our analytic formalism, $I^*$ is obtained by integrating Eqs. \eqref{eq:dSkldt} and \eqref{eq:dIkldt} until convergence is reached towards a stable manifold. Predictions of our analytic formalism are compared in Fig. \ref{fig:bif} with the outcome of Monte-Carlo simulations for static and adaptive networks with an initial Poisson degree distribution. For comparison, we also illustrate the analytic predictions of the low order approach of Gross \emph{et al.} (see Appendix). 

Figure~\ref{fig:bif_z20}, where each node has a mean degree of $\langle k \rangle = 20$, corresponds to the case treated previously by Gross \emph{et al.} in~\cite{gross06_prl}. In this highly connected limit, we see that both analytic formalisms are able to reproduce the correct equilibrium behavior of the system with and without rewiring. On Figs.~\ref{fig:bif_z7} and~\ref{fig:bif_z2}, we decrease the mean degree of the system to $\langle k \rangle = 7$ and 2 respectively. We see that our formalism continues to remain valid as $\langle k \rangle$ diminishes, while the formalism of Gross \emph{et al.} looses its accuracy. This result can be mainly explained by the fact that unlike in our approach, the equations of Gross \emph{et al.} do not distinguish between the individual behaviors of nodes with different degrees. These behaviors become increasingly heterogeneous as $\langle k \rangle$ is decreased~\footnote{For example, in the case where $\langle k \rangle = 20$, the behavior of a degree 18 node will not differ much from that of a degree 22 node. However, in the case where $\langle k \rangle = 2$, there will be a huge difference between the behavior of a degree 0 node and the behavior of a degree 4 node.}. Finally, in Fig.~\ref{fig:bif_z2}, we see that analytic predictions for $I^*$ are more accurate on adaptive than on static networks. This is due to the fact that link rewiring induces a certain amount of \emph{shuffling} in the network connections. In this case the history of the transmission events that did or did not happen becomes less important and the description of the system at a coarse-grained level is more accurate. 
 
For the systems with link rewiring shown in Fig.~\ref{fig:bif}, we can clearly see the existence of a bistable regime characterized by two first order transitions. To these discontinuous transitions correspond two thresholds: the \emph{persistence threshold} $\beta_{\textrm{per}}$, from which an already well-established epidemic can persist in the population, and the \emph{invasion threshold} $\beta_{\textrm{inv}}$, where the disease-free state become unstable for all finite values of $\epsilon$. These features have already been recognized as generic features of epidemic models on adaptive networks~\cite{gross06_prl, shaw08_pre, risau-gusman09_jtb}. 

After illustrating the effect that link rewiring has on the stationary disease prevalence in the system, we can also study how it affects the topology of the underlying network. The case of higher interest is the topology of the endemic state, where the system is in active equilibrium.

\begin{figure}[t]
\mbox{
\subfigure[\ $\gamma = 0$]{\includegraphics[width = .45\columnwidth]{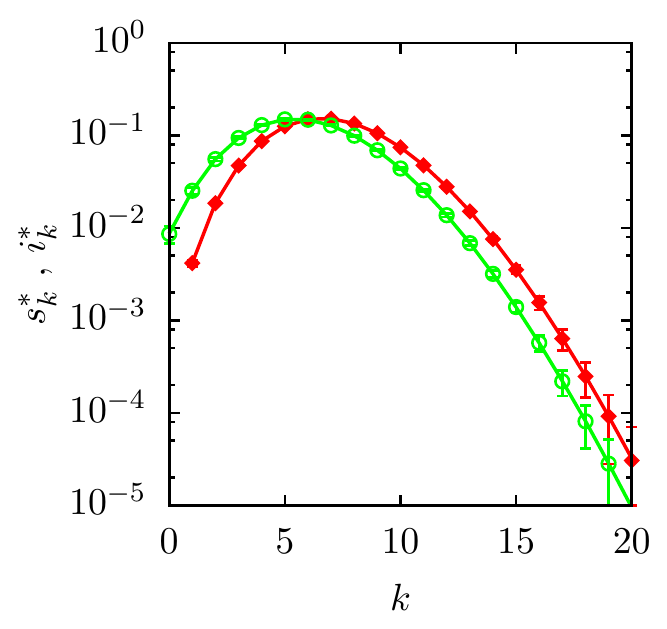} \label{fig:ddeq00}} 
\subfigure[\ $\gamma = 0.025$]{\includegraphics[width = .45\columnwidth]{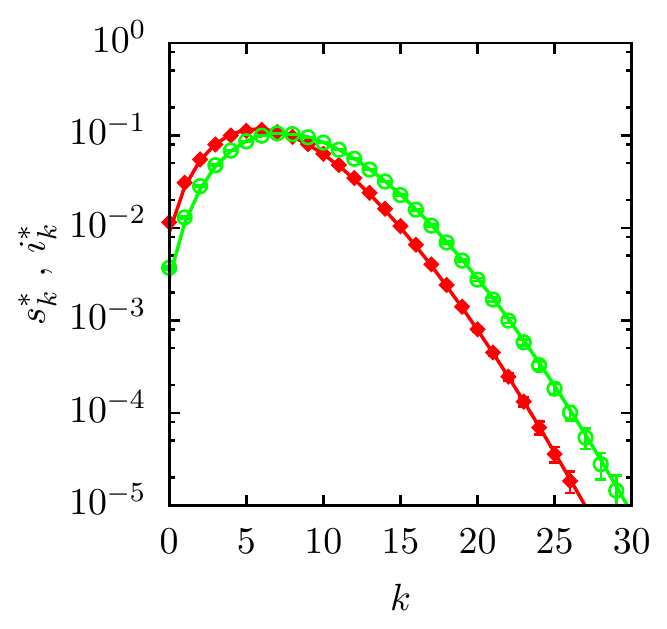} \label{fig:ddeq25}}} \\
\mbox{
\subfigure[\ $\gamma = 0.050$]{\includegraphics[width = .45\columnwidth]{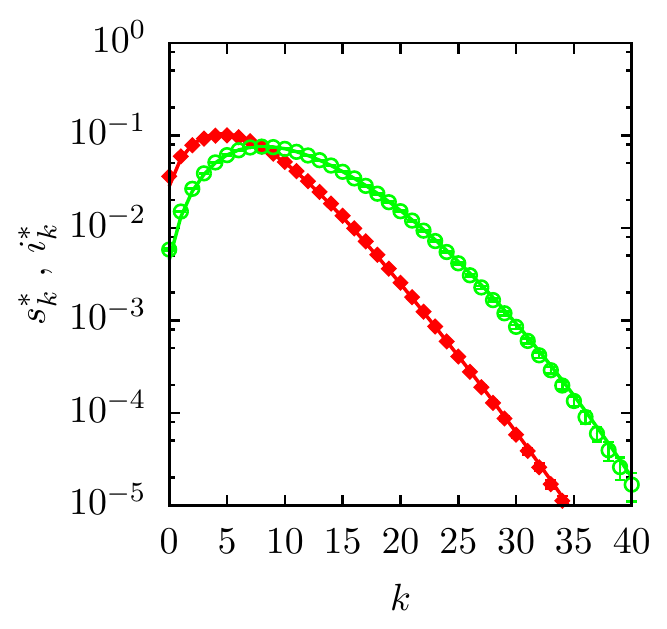} \label{fig:ddeq50}}   
\subfigure[]{\includegraphics[width = .45\columnwidth]{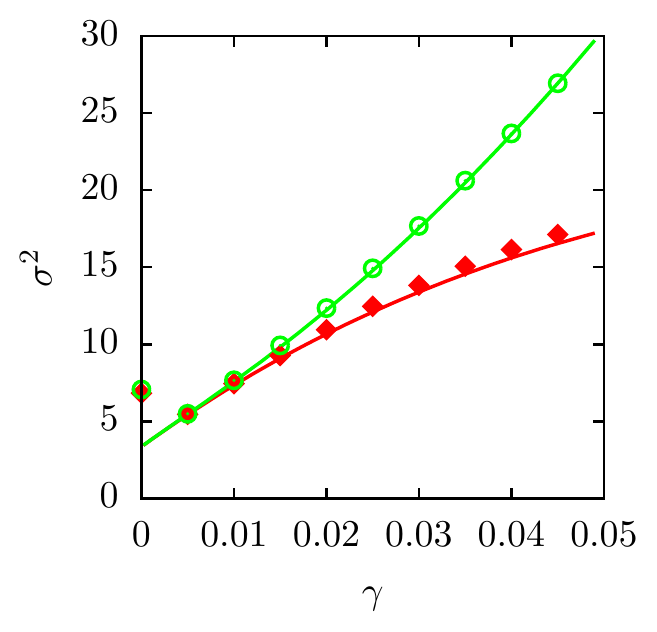} \label{fig:ddeqvar}}} 
\caption{(Color online) Normalized stationary degree distributions $s_k^*$ and $i_k^*$ in the endemic state for susceptible (open circles) and infectious (filled diamonds) nodes for $\alpha=0.005$, $\beta=0.008$, and (a) $\gamma = 0$, (b) $\gamma = 0.025$, and (c) $\gamma=0.050$. (d) Variance $\sigma^2$ of the stationary degree distributions of susceptible and infectious nodes versus the rewiring rate $\gamma$. The initial network used has a Poisson degree distribution with $\langle k \rangle=7$. Points and error bars (if larger than marker size) correspond to the mean and standard deviation computed over (a-c) 1000 and (d) 250 Monte-Carlo simulations; solid lines are the predictions of our analytic approach.  \label{fig:ddeq}}
\end{figure}

In Fig. \ref{fig:ddeq}, we illustrate the normalized stationary degree distributions observed at various rewiring rates in the endemic state of a system with an initial \emph{P} network and $\langle k \rangle = 7$. In our formalism, they are given by $s_k^* \equiv \sum_l S_{kl}^*/S^*$ for susceptible nodes and $i_k^* \equiv \sum_l I_{kl}^*/I^*$ for infectious nodes. For a static network, shown in Fig.~\ref{fig:ddeq00}, both stationary degree distributions follow a Poisson distribution. The peak of $i_k^*$ is found at higher degree than the peak of $s_k^*$ because high degree nodes are more likely to get infected. For adaptive networks, shown on Figs.~\ref{fig:ddeq25} and~\ref{fig:ddeq50}, both stationary degree distributions get significantly broader, particularly for susceptible nodes. In Fig. \ref{fig:ddeqvar}, the variance $\sigma^2$ of both distributions is plotted versus the rewiring rate $\gamma$. $\sigma^2$ is a smoothly increasing function of $\gamma$ for both distributions, its increasing rate being greater for susceptible nodes. For comparison, we also indicated the variance of the stationary distributions on a static network ($\gamma = 0$). Our results show no apparent continuous transition between the degree distributions at equilibrium on static and adaptive networks. Starting from the equilibrium topology in the adaptive regime, it is therefore impossible to recover the initial topology by slowly decreasing the rewiring rate of the system. We will return to the implication of this observation in more details in Sec.~\ref{sec:endemic}. 

The results presented in Fig.~\ref{fig:ddeq} confirm that link rewiring has two main effects of opposite epidemiological consequences~\cite{gross06_prl}. On the one hand, it locally promotes the isolation of infectious individuals, but on the other hand, it triggers the formation of highly connected individuals, which acts as \emph{superspreaders} of the disease. This dual effect may be responsible for the apparition of a bistable regime in parameter space, which is not observed in static networks. 

The effects of link rewiring on the topology of adaptive networks have been previously observed in stochastic simulations~\cite{gross06_prl, shaw08_pre}, but until now, no analytic approach was able to model them correctly. A previous attempt by Shaw and Schwartz~\cite{shaw08_pre} has been unsucessful, basically because their formalism was not able to account for the correlations between the state of neighboring nodes. The results shown in Fig. \ref{fig:ddeq} highlight that our improved compartmental formalism is able to capture with great accuracy the degree distributions of the system at equilibrium. By characterizing nodes by their total and infectious degree, we are able to overcome the correlation problems faced in~\cite{shaw08_pre}.

\subsection{Comparison of phase diagrams for different initial networks \label{sec:phase}}

\begin{figure}[t]
\centering
\includegraphics[width = 1.0\columnwidth]{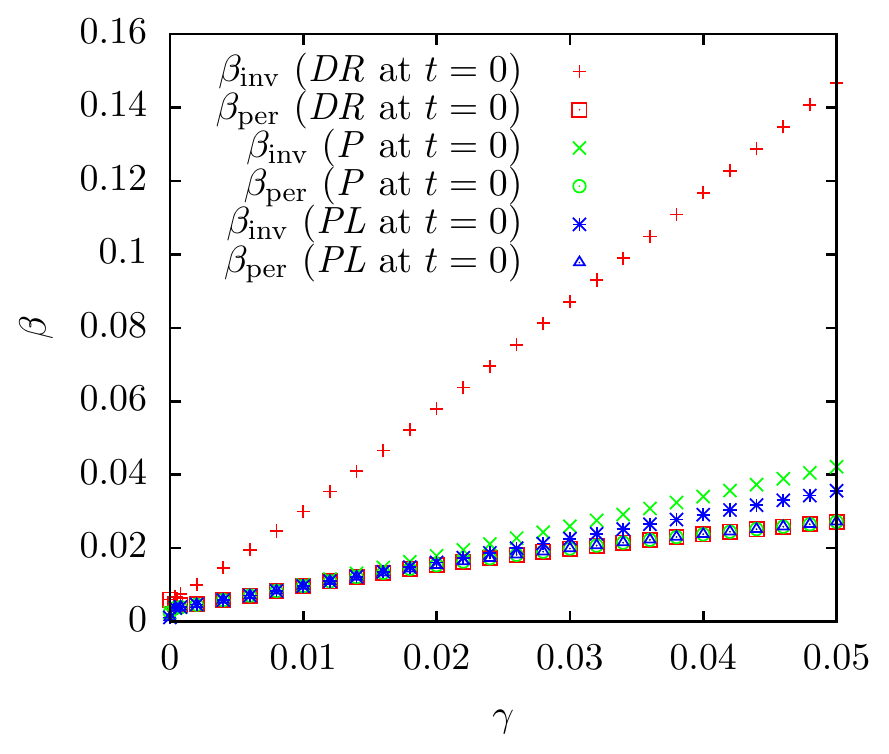} 
\caption{(Color online) Location of the persistence $\beta_{\textrm{per}}$ and invasion $\beta_{\textrm{inv}}$ thresholds versus the rewiring rate $\gamma$ as computed from our analytic approach at $\alpha=0.005$ for systems featuring the same mean degree $\langle k \rangle = 2$ but different initial degree distributions ($p_k^{DR}$, $p_k^P$ and $p_k^{PL}$). For each initial degree distribution, the upper-most region corresponds to the endemic state, the lower-most region to the disease-free state, and the inner region to the bistable regime. Except for very small values of $\gamma$, the persistence threshold of all three networks is the same (lowest curve). \label{fig:thresholds}}
\end{figure}

In the last subsection, we have studied the behavior of the model on adaptive networks with an initial Poisson degree distribution. For these particular initial networks, we have confirmed that there exists a bistable region at finite rewiring rate in parameter space. In this region, the initial disease prevalence plays an important role in determining if an epidemic will either die out or persist in the population. Moreover, we saw in Sec.~\ref{sec:timeevol} that the initial network topology also influences the evolution and outcome of a system. We now study the location of the persistence and invasion thresholds in systems featuring different initial topologies. 

Phase diagrams in the plane $(\gamma,\beta)$ for three systems featuring different degree distributions at the initial stage with $\langle k \rangle=2$ are illustrated in Fig.~\ref{fig:thresholds}. The location of both thresholds $\beta_{\textrm{per}}$ and $\beta_{\textrm{inv}}$ were obtained with our improved compartmental formalism using a bisecting algorithm, with $\epsilon=0.0001$ for $\beta_\textrm{inv}$ and $\epsilon=0.99$ for $\beta_\textrm{per}$. Figure~\ref{fig:thresholds} shows that all three networks display a bistability region between regions where only one stationary state, either endemic or disease-free, is stable. At fixed recovery rate $\alpha$, the extent of this bistability region depends on the rewiring rate $\gamma$ and the initial topology of the network. The invasion threshold $\beta_{\textrm{inv}}$ grows much faster as $\gamma$ is increased in systems with an initial \emph{DR} network than in systems with initial \emph{P} and \emph{PL} networks. These results suggest that for the same link density $\langle k \rangle$ and at small $\epsilon$, \emph{link rewiring as a disease control strategy is more efficient in homogeneous networks}, i.e. networks with small fluctuations in their degree distribution. As we mentioned previously, adaptive rewiring tend to supress disease propagation on a local scale, but has the potential to create high degree susceptible nodes on a global scale, which favors the spreading of the disease. On static networks, the initial spreading phase is known to be slower on homogeneous networks than on strongly heterogeneous networks~\cite{barthelemy04_prl}. When an adaptive rewiring rule is added, it is then easier for homogeneous networks to hinder the initial propagation of the disease on a local scale before it reaches a macroscopic prevalence and a critical concentration of high degree nodes is attained. Consequently, $\beta_{\textrm{inv}}$ is higher in homogeneous adaptive networks. 

Except at very small rewiring rates, Fig.~\ref{fig:thresholds} shows that the persistence threshold is the same for the three systems. Since this threshold marks the point from which a stable endemic state appears in the system, this supports the existence of a universal endemic state common to those systems, regardless of the initial topology. We will return to this point in Sec.~\ref{sec:endemic}. We believe that the persistence thresholds differ when $\gamma$ is small because the network does not evolve rapidly enough in all systems to converge towards a topology on which an endemic state would be stable, even if such a topology exists. 

The results presented in this section illustrate the importance of initial conditions in determining the global outcome, either endemic or disease-free, of a given epidemic scenario. The initial network topology determines the size of the bistability region, and inside the latter, the initial disease prevalence determines which stationary state will be reached.

\section{Further remarks on the endemic state \label{sec:endemic}}

In the last sections, we have gathered much evidence supporting the claim that the endemic state found in systems featuring an adaptive network is only determined by the dynamical parameters of the system, $\{\alpha,\beta,\gamma\}$, and the link density $\langle k \rangle$. For a given set of these parameters, the endemic state appears to be \emph{universal}, i.e. it does not depend on the particular initial conditions of the system. 

\begin{figure}[t]
\centering
\includegraphics[width = 0.8\columnwidth]{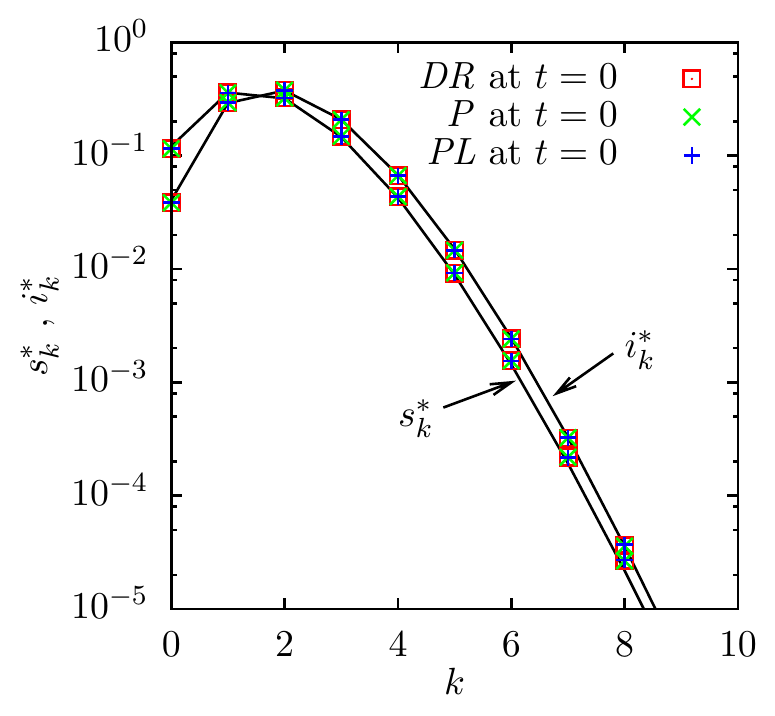} 
\caption{(Color online) Normalized stationary degree distributions $s_k^*$ and $i_k^*$ in the endemic state for susceptible and infectious nodes in three systems with different initial network topologies characterized by $\langle k \rangle = 2$. Simulation parameters are $\alpha=0.005$, $\beta = 0.006$, and $\gamma = 0.02$. Points correspond to the mean computed over $1000$ Monte-Carlo simulations; solid lines are the predictions of our analytic approach. \label{fig:ddeqend}}
\end{figure}

Let us briefly recall the results that corroborate this idea. For systems with the same value of $\{\alpha,\beta,\gamma\}$ and $\langle k \rangle$, we showed in Sec.~\ref{sec:timeevol} that even if their evolution towards the endemic state, if ever reached, is different, the disease prevalence $I$ converges in all cases to the same value. In addition, Fig.~\ref{fig:ddeqend} displays the fact that their degree distributions also converge to the same distribution. In Sec.~\ref{sec:bif}, we found no continuous transition in \emph{P} networks between the stationary degree distributions of susceptible and infectious individuals as link rewiring is turned on. Moreover, we computed in Sec.~\ref{sec:phase} the value of the persistence treshold $\beta_{\textrm{per}}$, and found that it has the same value in systems featuring adaptive networks with different initial topologies but the same mean degree $\langle k \rangle$. 

Is it reasonable to ask if this claim makes sense. In the coevolutionary model studied here, susceptible individuals are allowed to avoid contact with infectious individuals by changing acquaintances. As the disease propagates, links are being rewired and the network slowly loses memory of its initial structure. As already mentioned, the endemic state of the system is active. For a system able to reach it, the coevolution process between state and topology lasts for a infinitely long time. Hence, at some point, the information about its initial structure is completely lost.

This phenomenon can be interpreted in the framework of statistical mechanics. The endemic state can be thought as the state of maximum entropy of the system. It only depends on the \emph{density} parameter of the system, $\langle k \rangle$, and the \emph{interaction} parameters between \emph{particles} (nodes) of the system, $\{\alpha,\beta,\gamma\}$. As the system evolves dynamically towards the state of maximum entropy, information is lost. Therefore, the evolution process towards the endemic state is \emph{irreversible}.
 
However, even if there is much evidence in favor of the existence of a universal endemic state for given $\{ \alpha, \beta, \gamma\}$ and $\langle k \rangle$, this statement still remains at the conjecture level. Since our improved compartmental formalism does not seem to admit an analytic solution for the equilibria of the system, it is impossible to mathematically demonstrate that the solution for the endemic state does not depend on the particular initial conditions of the system. Another approach may be needed to solve this problem.

\section{Conclusion \label{sec:conclusion}}

In the spirit of the formalism presented in Appendix of~\cite{noel09_pre}, we have introduced an improved compartmental formalism in the framework of a simple SIS model on networks featuring an adaptive rewiring rule~\cite{gross06_prl}. In our approach, individuals are put in compartments according to their state of infectiousness, their \emph{total degree} $k$ and their \emph{infectious degree} $l$. With these considerations, a set of ODEs describing the dynamics of the system is obtained and can be integrated numerically to yield its evolution and stationary states. Theoretical predictions were found to be in excellent agreement with numerical results for adaptive networks with various degrees of heterogeneity at the initial stage. Being the first capable of reproducing the complete time evolution of both dynamical elements, \emph{process} and \emph{structure}, the approach presented in this paper marks an important step forward in understanding the complex behavior of adaptive networks. 

As a pedagogical example, we have analyzed in details the coevolution of disease and topology in a system featuring a degree-regular network at the initial stage. By tracking the evolution of meaningful observables, we were able to point out the dual effect of link rewiring in the population. Beside bringing better insights about the interplay between disease and topology on adaptive networks, this simple example showed that our formalism is very well-suited for the study of these complex systems.

Moreover, the results obtained show that the initial conditions, i.e. disease prevalence and network topology, play an important role in determining the evolution and outcome of a particular epidemic scenario on an adaptive network. It does not only affect the speed at which stationarity is reached, but can also determine \emph{which} stationary state is reached - either endemic or disease-free. In contrast, the properties of the endemic state do not seem to be affected by the initial topology of the network. We have presented strong evidence that it only depends on the dynamical parameters of the system, $\{\alpha,\beta,\gamma\}$, and the mean degree of the network, $\langle k \rangle$. However, since our model cannot be solved analytically, this conjecture remains to be proved. 

The use of the model of Gross \emph{et al.} as the framework of this paper served the purpose of \emph{proof of concept}. Despite its appealing simplicity, this particular model is actually lacking in realism. Possible directions for further research could consist of including more realistic features in epidemic models on adaptive networks. A first step in this direction could be the introduction of cliques~\cite{hebert10_arxiv} in the network to account for the community structure observed in many real-world networks~\cite{girvan02_pnas}. Another interesting effect to implement would be some mechanism of \emph{preferential rewiring}. For example, this could be modeled by choosing nodes with a probability proportionnal to the inverse of their degree in the rewiring process. This would account for the fact that people may be aware that being in contact with an highly-connected individual is more dangerous than with someone having only few acquaintances. This feature could potentially hinder to some extent the formation of high-degree susceptible nodes in the network.

\begin{acknowledgments}
Our research team is grateful to NSERC (VM and LJD), FQRNT (VM, LHD, and LJD), and CIHR (PAN, LHD, and AA) for financial support. We also acknowledge Thilo Gross for helpful discussions.
\end{acknowledgments}

\appendix*

\section{The formalism of Gross \emph{et al.}}

Until now, existing models of epidemic spreading on adaptive networks have been studied analytically with the help of low order moment-closure approximations~\cite{gross06_prl, zanette08_jbp, shaw08_pre, risau-gusman09_jtb}. Here we present the formalism of Gross \emph{et al.} for SIS dynamics on adaptive networks~\cite{gross06_prl}. 

Let $[X_1]$, $[X_1 X_2]$ and $[X_1 X_2 X_3]$, where $X_i \in \{S,I\}$, represent the zeroth, first and second order moments of the system. $[X_1]$ correspond to the fraction of $X_1$ nodes in the network, $[X_1 X_2]$ is the density of $X_1 X_2$ links per node and $[X_1 X_2 X_3]$ is the density of $X_1 X_2 X_3$ triplets per node in the networks. The variables used by Gross \emph{et al.} then relate to those in our formalism by the relations $[S]=S$, $[I]=I$, $[SS]=S_S/2$, $[II]=I_I/2$, $[SI]=S_I=I_S$, $[SSI]=S_{SI}$, and $[ISI]=S_{II}$. Conservation relations \eqref{eq:c1} and \eqref{eq:c2} can now be written as $[S] + [I] = 1$ and $[SS] + [SI] + [II] = \langle k \rangle/2$. In addition to the latter constraints, the dynamics of the zeroth and first order moments of the system are described by the following balance equations:
\begin{align}
\frac{d[I]}{dt} &= \beta [SI] - \alpha [I] \ ,\label{eq:dIdt} \\ 
\frac{d[II]}{dt} &= \beta ([SI] + [ISI]) - 2\alpha[II] \ , \label{eq:dIIdt} \\
\frac{d[SS]}{dt} &= (\alpha+\gamma)[SI] - \beta[SSI] \ . \label{eq:dSSdt}
\end{align}
This dynamical system captures the effect of rewiring via the first term in \eqref{eq:dSSdt}. However, it does not yet represent a closed model because of the appearance of the second order moments in the last two equations. For this reason, the \emph{pair approximation} technique is used. The latter consists of approximating the second order moments by $[X_1X_2 X_3]\approx[X_1X_2][X_2X_3]/[X_2]$, which gives in this case $[ISI] \approx [SI]^2/[S]$ and $[SSI]\approx 2[SS][SI]/[S]$. Together with this approximation, Eqs. \eqref{eq:dIdt}-\eqref{eq:dSSdt} and the two conservation relations constitute a closed model which can be studied in the framework of nonlinear dynamics.

The main difference between our improved compartmental formalism and previous approaches resides in the level of coarse-graining in the system. The variables $S_{kl}$ and $I_{kl}$, on which our formalism is based, simply correspond to the underlying distributions of the moments $[X_1]$ and $[X_1 X_2]$.


\end{document}